\documentclass[conference]{IEEEtran}
\IEEEoverridecommandlockouts
\usepackage{cite}
\usepackage{amsmath,amssymb,amsfonts}
\usepackage{algorithmic}
\usepackage{comment}
\usepackage{graphicx}
\usepackage{soul}
\usepackage{tabularx}
\usepackage{array}
\usepackage{textcomp}
\usepackage{xcolor}
\usepackage{tcolorbox}
\usepackage{caption}
\usepackage{threeparttable}
\usepackage{placeins}
\usepackage{orcidlink}
\usepackage{booktabs}
\usepackage{gensymb}
\usepackage{hyperref}
\usepackage{siunitx}
\usepackage{url}
\def\BibTeX{{\rm B\kern-.05em{\sc i\kern-.025em b}\kern-.08em
    T\kern-.1667em\lower.7ex\hbox{E}\kern-.125emX}}
\begin{document}

\title{CNN Models for Microphone Array Covariance Matrix Upsampling and Acoustic Imaging\\
\thanks{The authors wish to acknowledge CSC—IT Center for Science, Finland, for computational resources.} %Identify applicable funding agency here. If none, delete this.
}

\author{
\IEEEauthorblockN{
Marianthi Adamopoulou\,\orcidlink{0000-0001-7410-0483}\IEEEauthorrefmark{1},
Parthasaarathy Sudarsanam\,\orcidlink{0009-0009-3751-6469}\IEEEauthorrefmark{2},
David Diaz-Guerra\,\orcidlink{0000-0002-1041-0498}\IEEEauthorrefmark{2},
Meng Jiang\,\orcidlink{0000-0002-8253-7535}\IEEEauthorrefmark{1},\\
Archontis Politis\,\orcidlink{0000-0002-0595-2356}\IEEEauthorrefmark{2},
Seyed Jalaleddin Mousavirad\,\orcidlink{0000-0001-8661-7578}\IEEEauthorrefmark{1},
Tuomas Virtanen\,\orcidlink{0000-0002-4604-9729}\IEEEauthorrefmark{2},
Jan Lundgren\,\orcidlink{0000-0003-1819-6200}\IEEEauthorrefmark{1},
}
\IEEEauthorblockA{
\IEEEauthorrefmark{1}Dept. of Comp. and Elec. Engineering Mid Sweden University
Sundsvall, Sweden
}

\IEEEauthorblockA{
\IEEEauthorrefmark{2}Audio Research Group, Tampere University Tampere, Finland
}

\IEEEauthorblockA{
\IEEEauthorrefmark{1}Email: \{name.lastname\}@miun.se
}
\IEEEauthorblockA{
\IEEEauthorrefmark{2}Email: \{name.lastname\}@tuni.fi
}
}

\begin{comment}
\author{\IEEEauthorblockN{1\textsuperscript{st} Given Name Surname}
\IEEEauthorblockA{\textit{dept. name of organization (of Aff.)} \\
\textit{name of organization (of Aff.)}\\
City, Country \\
email address or ORCID}
\and
\IEEEauthorblockN{2\textsuperscript{nd} Given Name Surname}
\IEEEauthorblockA{\textit{dept. name of organization (of Aff.)} \\
\textit{name of organization (of Aff.)}\\
City, Country \\
email address or ORCID}
\and
\IEEEauthorblockN{3\textsuperscript{rd} Given Name Surname}
\IEEEauthorblockA{\textit{dept. name of organization (of Aff.)} \\
\textit{name of organization (of Aff.)}\\
City, Country \\
email address or ORCID}
\and
\IEEEauthorblockN{4\textsuperscript{th} Given Name Surname}
\IEEEauthorblockA{\textit{dept. name of organization (of Aff.)} \\
\textit{name of organization (of Aff.)}\\
City, Country \\
email address or ORCID}
\and
\IEEEauthorblockN{5\textsuperscript{th} Given Name Surname}
\IEEEauthorblockA{\textit{dept. name of organization (of Aff.)} \\
\textit{name of organization (of Aff.)}\\
City, Country \\
email address or ORCID}
\and
\IEEEauthorblockN{6\textsuperscript{th} Given Name Surname}
\IEEEauthorblockA{\textit{dept. name of organization (of Aff.)} \\
\textit{name of organization (of Aff.)}\\
City, Country \\
email address or ORCID}
}
\end{comment}

\maketitle

\begin{abstract}
Acoustic imaging visualization is a core methodology in acoustics, enabling spatial analysis of sound sources and acoustic scenes. %Sensor limitations arise in real-world applications, motivating the adoption of deep learning methods to improve system performance by minimizing hardware complexity. 
However, limited sensor availability in practical systems motivate approaches that enhance spatial resolution without increasing the hardware complexity. In this paper, we focus on upsampling virtually a tetrahedral 4-microphone array to a spherical 32-microphone array by estimating the covariance matrices of the channels employing deep learning techniques. Five neural network architectures are investigated for covariance upsampling for acoustic imaging using the real-world STARSS23 dataset. % In this paper, five deep learning model architectures are developed and compared for spatial upsampling employing the STARSS23 real-world dataset. 
These models are developed to estimate a 32-microphone, time–frequency covariance matrix from a 4-microphone input covariance representation.
%The input of the model is 4D 4-channel covariance matrix predicting 4D 32-channel covariance matrix time-frequency dependent. 
%The models are constructed with 2D convolution layers to handle the complexity of covariance matrices. In addition, Frequency Dynamic Convolution layers are employed to take into consideration the characteristics of frequency dependency of the matrices. 
The proposed architectures are based on 2D convolutional layers to capture the underlying spatial–spectral structure of covariance matrices, and are further enhanced with frequency dynamic convolution to model their frequency-dependent properties. The proposed architectures are evaluated in terms of root mean square error (RMSE) and using delay-and-sum beamforming acoustic imaging. Quantitative results show that all models outperform a random-guess baseline, which yields an RMSE of 0.548, with the best-performing architecture achieving an RMSE of 0.432. We analyze qualitatively the performance of the proposed models through beamforming heatmap visualizations derived from the 4-channel input covariance, the 32-channel ground truth, and the predicted 32-channel covariance matrices. These results demonstrate that covariance upsampling significantly enhances the effective performance of the 4-channel microphone array, producing sound maps that closely resemble those obtained with the 32-channel array.
%RMSE values and delay-and-sum beamforming are computed to evaluate the proposed architectures. Heatmaps for both 4-channel input data, 32-channel ground truth, and 32-channel prediction are presented for model evaluation. Across all frequency bins, the proposed deep learning architectures outperform random guesses, improving the performance of the 4-channel microphone array, producing sound maps close to the 32-channel ground truth, yet more focused of the sound sources and cleared from reflections.
\end{abstract}


\section{Introduction}\label{section: introduction}

Acoustic imaging visualizes the spatial distribution of sound sources in an environment by exploiting spatial information captured by microphone arrays followed by multichannel signal processing techniques. It is a key technology in instrumentation related to identification, characterization, and localization of sound sources in noisy and reverberant conditions.
It is widely used in automotive and aerospace industries \cite{rittenschober2024high, merino2019review}, marine surveillance \cite{dantzker2025deciphering}, monitoring of wind turbines \cite{ding2023acoustic}, pipe leakages or electrical line discharges \cite{pihera2020partial}, and industrial machine monitoring or fault diagnosis \cite{orman2013usage, xu2024investigation, fiebig2020use}.

Despite their effectiveness, acoustic imaging systems are fundamentally constrained by the size, geometry, and number of microphones in the array. Achieving high spatial resolution requires large-size dense arrays with many sensors, with increased hardware costs and computational complexity. In practical systems, such as mobile or embedded platforms, these requirements limit the scalability of the array. As a result, acoustic imaging with sparse or compact arrays often exhibits reduced spatial resolution and degraded localization performance, motivating methods that enhance spatial representations without increasing the physical array size and the number of microphones \cite{lubeck, 10694489}. Super-resolution approaches have been proposed that can produce acoustic images beyond the natural spatial resolution of the array. Those can be deconvolution-based \cite{brooks2006deconvolution}, parametric \cite{schmidt1986multiple, dougherty2014functional}, spatial re-assignment based \cite{mccormack2019sharpening}, or based on sparse recovery \cite{epain2009application}. Many of those methods rely on statistical dependencies between the microphones of the array, captured in the spatial covariance matrix (SCM) of the array signals. 
%[Spatial Upsampling of Sparse Spherical Microphone Array Signals].

%Conventional upsampling methods (interpolation?) and their problems (not robust under noise and reverb?)...In contrast, deep learning offers a data-driven alternative capable of learning complex spatial relationships directly from multichannel observations, making it well suited for enhancing spatial representations in constrained acoustic imaging scenarios.

 In this work, we focus on a similar super-resolution task applied to spherical microphone arrays (SMAs), which offer a fully spherical field-of-view and uniform imaging resolution at all directions. Therefore, to address hardware limitations, a 4-channel microphone array along with deep-learning are employed to virtually upsample the array into a 32-channel SMA, by predicting the covariance matrices of the missing channels, aiming to improve the performance of the system. Acoustical imaging with SMAs is often performed in the spherical harmonic (SH) transform domain due to theoretical and computational advantages. Upsampling the SH transformed microphone signals to harmonic orders higher than the ones provided naturally by the SMA is, then, one way to spherical super-resolution imaging.
%This can be done by estimating different kind of parameters that would be obtained with an array with a high number of microphones from the signals obtained with an array with a lower number of microphones. 
%For example, 
In \cite{salvador2018enhancing, lubeck2023spatial}, virtual microphone signals are generated by interpolating the real ones, increasing artificially the SH transform order of the array. In \cite{epain2009application, wabnitz2011upscaling} low-resolution SH signals are upscaled directly to higher-order ones by assuming spatial sparsity on the sound sources, while \cite{routray2019deep, chatzimoustafahigher} train deep-learning models for the same upscaling task.

An alternative approach to upscaling or interpolating the array signals is to instead try to predict the interchannel signal statistics of a larger or denser virtual microphone array, in the form of its SCM, from the real observed SCM. Standard normal or super-resolution imaging methods can be then applied to such an upscaled SCM. 
%However, for sound source localization (SSL) or acoustic imaging applications, the inter-channel covariance matrix of the array contains all the information needed. 
In \cite{roman2024}, the deep-learning image super-resolution model Deep Back Projection Network (DBPN) \cite{DBPN2018} is applied on the SCM of a 4-microphone tetrahedral array to estimate the SCM of a 32-microphone SMA, which is used to generate acoustic images using DeepWave, a deep-learning model that generates acoustical images from covariance matrices \cite{simeoni2019deepwave}. The replacement of DeepWave by a self-supervised model is later proposed in \cite{LatentAcoustic}, but the upsampling of the SCM is still done with DBPN. 

Even if both can be seen as 2D signals, covariance matrices are very different from the images DBPN was designed for. In images, the spatial distance between all the adjacent pixels is supposed to be the same, which is not the case for inter-channel covariance matrices, where the distance between elements depend on the array geometry and the order in which the microphones have been numbered. This means that using 2D convolutions over the covariance matrices, as DBPN does when applied to them, might not be and optimal choice. In addition, if DBPN is applied independently to the covariance matrices of each time-frequency bin as done in \cite{roman2024, LatentAcoustic}, the model cannot exploit the correlations between different bins.

In the present work, we propose a deep learning framework for SCM upsampling that operates directly on time–frequency dependent SCMs, preserving both temporal and frequency-specific spatial structure. As in \cite{roman2024, LatentAcoustic}, our method maps low-dimensional SCM representations from sparse microphone arrays to high-dimensional SCMs, enabling general-purpose acoustic imaging. However, we use models specifically designed for covariance matrices: 2D convolutional layers are applied across the time and frequency dimensions and the real and imaginary parts of the non-redundant elements of the matrices are treated as convolution channels. We train and evaluate the proposed approach on the real-world STARSS23 dataset and conduct a systematic comparison of convolutional neural network architectures, including frequency-dependent convolutional layers, to analyze their impact on SCM reconstruction performance. 

\begin{comment}
    The remainder of the paper is organized as follows. Section \ref{section: methodology} presents our proposed deep learning methodology. Sections \ref{section: evaluation} and \ref{section: results} describe the datasets used for training and evaluation and report the corresponding results. Finally, Section \ref{section: conclusion} summarizes our findings and discusses limitations and directions for future work.
\end{comment}

% The remainder of the paper is organized as follows. Section \ref{section: methodology} presents our proposed deep learning methodology and the datasets used for training and evaluation. Section \ref{section: results} report the corresponding results. Finally, Section \ref{section: conclusion} summarizes our findings and discusses limitations and directions for future work.

\section{Methodology}\label{section: methodology}
%In this work, we aim to upsample the spatial information captured by a sparse 4-channel subset of an EigenMike array to the corresponding full 32-channel spherical array. This is done by learning to upsample the complex SCM derived from the sparse array. We systematically study and compare five different convolutional neural network (CNN) architectures that operate directly in the covariance domain. Contrary to DBPN or other CNNs designed for image processing, we use the convolutions to incorporate time-frequency context to the computation of every covariance matrix, with the matrix itself not being treated as an image but as convolution channels. Therefore, we do not impose any prior assumptions about the array geometry into the model.

%Upsampling from low-resolution data to high-resolution with the upgrade of the size of the system virtually is challenging but carries many advantages over hardware complexity. In this work, we study and compare five different convolutional neural network (CNN) architectures that operate directly in the covariance domain to predict the 32-channel covariance matrix from the 4-channel covariance matrix that consist the input of the models. We use the convolutions to incorporate time-frequency context to the computation of every covariance matrix, with the matrix itself not being treated as an image but as convolution channels. Therefore, we do not impose any prior assumptions about the array geometry into the model.

In this work, we study and compare five different convolutional neural networks (CNN) to upsample the SCM of a microphone array with a reduced number of microphones into the one that would have been obtained with an array with more microphones. Specifically, we tackle the case of upsampling the SCM of a tetrahedral array of 4 microphones to the one obtained by a 32-microphone spherical array, though the same methodology could be used with different arrays. We use the convolutions to incorporate time-frequency context to the computation of every covariance matrix, with the matrix itself not being treated as an image but as convolution channels. Therefore, we do not impose any prior assumptions about the array geometry into the model.

%\textcolor{blue}{CNN architecture is chosen as the more suitable network for this case, as CNN's strength on audio signals equip the network with the necessary tools for precise predictions. It is known that CNNs outperform other approaches by learning complex relations in acoustics from raw data [A Comprehensive Review of Deep Learning,A Comprehensive Review of Deep Learning].}% Furthermore, CNN models are good in predicting statistical parameters and finding the relations between statistical weights, therefore a matrix such as covariance matrix as input is an ideal form of data for CNN architectures.[]}

\subsection{Input representation}

Let $X_m(t,f)$ be the short-time Fourier transform (STFT) of the signal captured by the $m$-th microphone of the array, which we compute using a Hann window with an FFT length of 512 samples, a 50\% hop size, and a sampling rate of 24 kHz. For both the 4-microhpone and the 32-microphone arrays, we can construct the instantaneous SCM of each time-frequency bin as
\begin{equation}
\mathbf{C}_x(t,f) = \mathbf{X}(t,f) \mathbf{X}^H(t,f)
\end{equation}
where $\mathbf{X}(t,f) = [X_1(t,f), \ldots, X_M(t,f)]^\top$ is the vector of STFT coefficients across the $M$ microphones, and $(\cdot)^H$ denotes the Hermitian transpose. These instantaneous SCMs capture the inter-channel spatial structure of the sound field at each time-frequency bin. To reduce noise and smooth the covariance matrix, the instantaneous SCMs are averaged over short temporal windows, producing 5 SCMs per second. Finally, we normalize the SCM of each time-frequency bin by dividing it by the value of the first element of its diagonal (i.e., the energy in the first sensor at that time-frequency bin) for an easier processing by the models. %This results in a time-varying, time–frequency dependent representation. 

%Since the SCMs are Hermitian, only the unique real components (including the diagonal) and unique imaginary components (excluding the diagonal) are retained. These components are stacked to form the input tensor $\mathbb{R}^{C \times F \times T}$, where $C=M^2$ is the number of unique covariance channels, $F$ is the number of frequency bins, and $T$ is the number of time frames. This tensor serves as the input to the neural network. In this work, 4-channel and 32-channel ground truth are presented accordingly for direct evaluation. %The network is trained to upsample this low-order covariance matrices to the corresponding high-order covariance matrices, which represent the spatial structure that would be captured by a dense spherical microphone array. Training is performed in a supervised learning setting by minimizing the mean squared error (MSE) between the predicted and target covariance matrices.

Since the SCMs are Hermitian matrices with real-valued diagonal, they only contain only $M^2$ non-reductant values: the real part of the diagonal elements and the real and the imaginary parts of the elements bellow the diagonal. These components are stacked as convolution channels to form a $\mathbb{R}^{C \times F \times T}$ tensor, where $C=M^2$ is the number of non-reductant covariance channels, $F$ is the number of frequency bins, and $T$ is the number of time frames. This tensor, computed from the signals of the tetrahedral and the spherical array, serves as the input and groundtruth output of the model respectively.

\subsection{Neural Network Architectures}\label{Arch}
\begin{figure*}
  \centering
  \includegraphics[width=\textwidth]{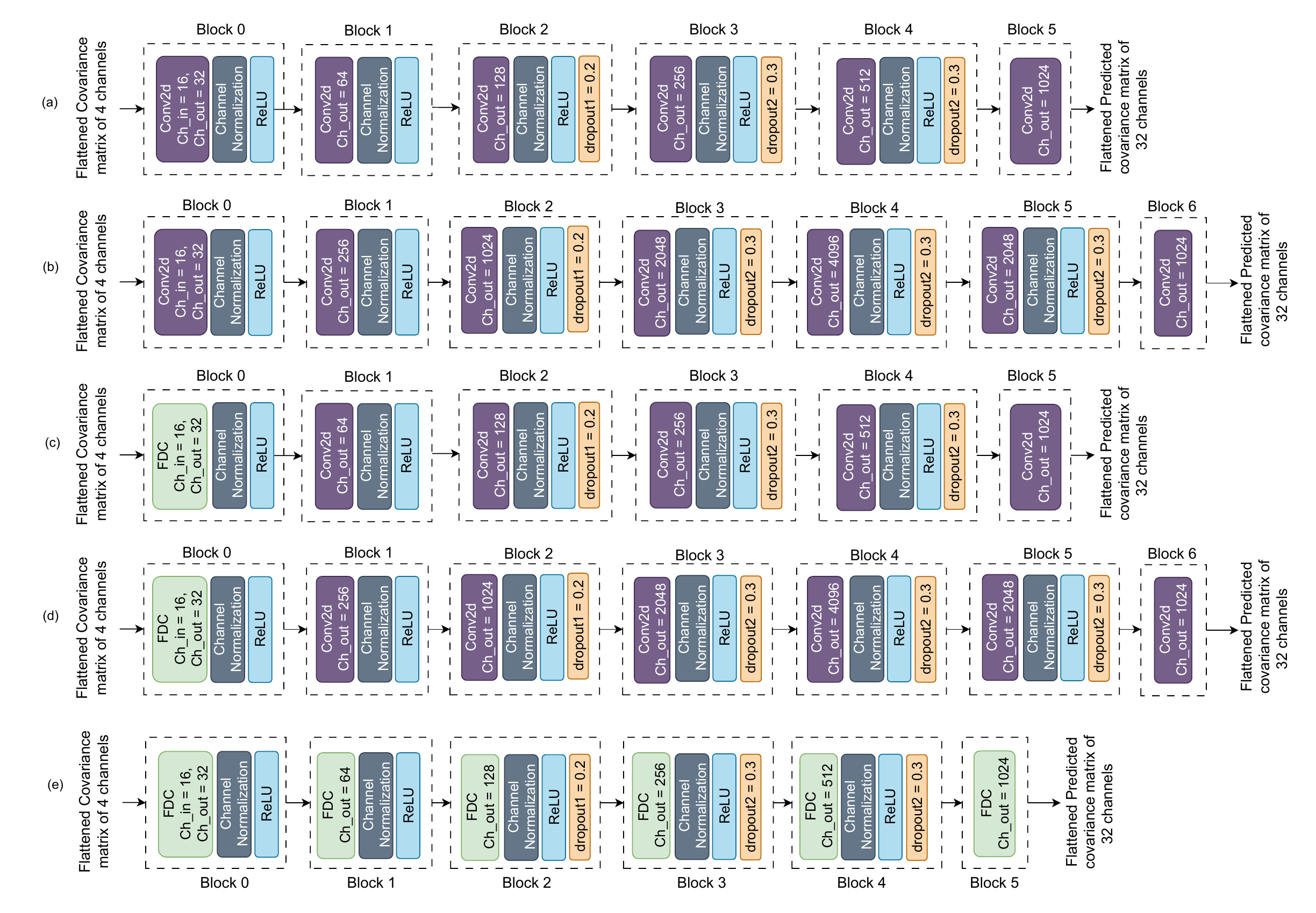}
  \caption{model architectures. (a) Base CNN, (b) Expanded CNN, (c) Hybrid FDC-CNN Base, (d) Hybrid FDC-CNN Expanded, (e) Full FDC-CNN}
  \label{fig:modelarch}
\end{figure*}

We investigate the CNN architectures for upsampling spatial covariance representations, as illustrated in Fig.~\ref{fig:modelarch}. All models are designed to map a low-dimensional covariance representation obtained from a sparse 4-channel array to a high-dimensional representation corresponding to a 32-channel array. They are trained in a supervised learning setting by minimizing the mean squared error (MSE) between the predicted and target SCMs.
\footnote{All employed codes are publicly available in: \url{https://github.com/marianthiadm/Upsampling-sparse-microphone-array-with-CNN}}
\paragraph{\textbf{Base CNN}}
The base CNN is composed of a sequence of six 2D convolutional blocks with kernel size $3 \times 3$ and progressively increasing channel dimensionality as shown in Figure \ref{fig:modelarch} (a). Starting from 16 input channels, corresponding to the flattened 4-channel covariance representation, the network gradually increases the number of channels until reaching 1024 output channels, corresponding to the $32$-channel covariance matrix. Each convolutional block consists of a 2D conv layer, which is employed for spectro-temporal analysis of the data, followed by channel normalization, to accelerate the training, and a ReLU activation function, while dropout is applied in the deeper layers to reduce overfitting. 

\paragraph{\textbf{Expanded CNN}}
The expanded CNN architecture extends the base CNN by increasing the representational capacity of the network as shown in Figure \ref{fig:modelarch} (b). It follows the same overall structure and layer sequence as the base CNN, but it introduces an additional channel expansion stages in which the channel dimension is increased beyond the target size. Specifically, the network expands the channel dimension up to 4096 channels before projecting it back to the required 1024 output channels. This overparameterization is intended to allow the network to learn richer intermediate spatial representations before reconstruction.

\paragraph{\textbf{Hybrid FDC-CNN Base}}
The previous models use the same convolutional kernels to process all the frequency bins, which might not be optimal considering that the SCMs are frequency dependent. To allow the models to process each frequency differently, we study the use of frequency dependent convolutions (FDC) implemented as 1D convolutions across time with a different kernels for each frequency bin. In a similar way to \cite{heikkinen2024neural}, the Hybrid FDC-CNN base model is equal to the base CNN just replacing the first convolutional layer by a FDC as shown in Figure \ref{fig:modelarch} (c).

%This architecture used frequency-aware processing by introducing Frequency Dynamic Convolution (FDC) layers as shown in Figure \ref{fig:modelarch} (c). Similar to the baseline CNN, it consists of six convolutional blocks for channel upsampling. However, in the first block, the standard CNN layer is replaced with FDC layer. This layer adapt their convolutional kernels across frequency bins, allowing the model to capture frequency-dependent spatial characteristics of the covariance representation. The remaining blocks use standard convolutional layers with channel normalization, ReLU activation, and dropout as the baseline.

\paragraph{\textbf{Hybrid FDC-CNN Expanded}}
This architecture also uses an FDC layer as shown in Figure \ref{fig:modelarch} (d). Similar to the Expanded CNN, it follows the exact same architecture, where the only difference is found in the first block. Thus, here as earlier, the standard CNN layer of the first block is replaced with an FDC layer. Contentiously, the remaining blocks use standard convolutional layers with channel normalization, ReLU activation, and dropout as the expanded model.

\begin{comment}
\paragraph{\textbf{Hybrid FDC-CNN}}
This architecture employed also FDC layers as shown in Figure \ref{fig:modelarch} (e), similar to the baseline CNN. Whereas, in this case FDC layers are applied in the first and last blocks. The remaining blocks use standard convolutional layers with channel normalization, ReLU activation, and dropout as the baseline.
\end{comment}

\paragraph{\textbf{Full FDC-CNN}}
The full FDC-CNN architecture further extends the frequency-aware design by employing FDC layers in all the convolutional blocks as shown in Figure \ref{fig:modelarch} (e). This model is the most computationally demanding among the proposed architectures, as every layer dynamically adapts its kernels to frequency-specific characteristics. Consequently, the extension was implemented on the base architecture, as the expanded model encompasses high complexity and training time. This fully frequency-adaptive design aims to maximize the model’s ability to capture fine-grained spectral–spatial dependencies in the covariance domain.

\begin{figure*}
  \centering
  \includegraphics[width=\textwidth]{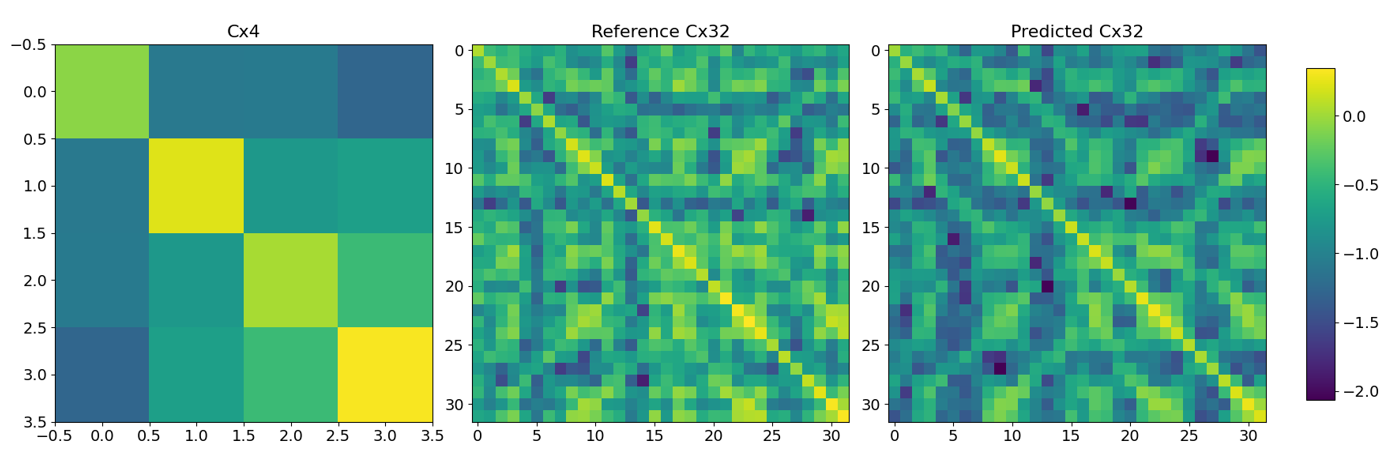}
  \caption{Frequency average of the SCMs for a single time frame. Left is the 4-channel input SCM, in the middle is the ground truth of the 32-channel SCM, and right is the estimation of the Hybrid FDC-CNN Expanded model.}
  \label{fig:covmat}
\end{figure*}

%\FloatBarrier
\subsection{Dataset}\label{section: dataset}
All models were trained on the development–train split of the Sony-TAu Realistic Spatial Soundscapes 2023 (STARSS23) dataset \cite{shimada2023starss23}, provided as part of DCASE 2024 Task 3. The dataset consists of multichannel recordings of real acoustic scenes captured in a variety of rooms and environments using a 32-channel EigenMike microphone array. The STARSS23 development dataset contains approximately 7.5 hours of recordings collected across 16 different rooms and includes 13 sound event classes, such as female and male speech, clapping, telephone, laughter, domestic sounds, footsteps, doors, music, musical instruments, water taps, bells, and knocking sounds \cite{David_g}. For the purposes of this study, the dataset has been segmented into 5-second clips to form the input of the models. It should be noted that certain segmented audio clips were affected by saturation, while others exhibited substantially higher energy over one channel; to avoid bias these segments were removed.

Model evaluation was performed on the STARSS23 public test dataset, which was not used for early stopping nor did it influence the training result in any way. The 4 channels that are used for the model input have been selected from the 32-channel EigenMike spherical array to approximate a tetrahedral array and correspond to those available in the public version of the dataset \cite{David_g}. %based on the first order ambisonic (FOA) rules. They are carrefully chosen to recreate the closest possible a tetrahedron microphone array following its transfer function formulas. More precisely, the mathematical equations behind this choice:

%\begin{equation}
%\label{eq:foa}
%\begin{aligned}
%H_1 (\varphi,\theta, f) &= 1 \\
%H_2 (\varphi,\theta, f) &= sin(\varphi) \ast cos(\theta) \\
%H_3 (\varphi,\theta, f) &= sin(\theta) \\
%H_4 (\varphi,\theta, f) &= cos(\varphi) \ast cos(\theta)
%\end{aligned}
%\end{equation}

% Therefore, the positions of the selected microphones that form the tetrahedral low-resolution input dataset are [DCASE challeng task 3 2024]: 
% \begin{equation}
% \label{eq:tetra}
% \begin{aligned}
% M_m:& (\varphi, \theta, r) \\
% M_1:& (45\degree, 35\degree, 4,2 cm) \\
% M_2:& (-45\degree, -35\degree, 4,2 cm) \\
% M_3:& (135\degree, -35\degree, 4,2 cm) \\
% M_4:& (-135\degree, 35\degree, 4,2 cm)
% \end{aligned}
% \end{equation}

\subsection{Training}\label{section: training}
All models are trained for 100 epochs using the Adam optimizer with a learning rate of $1\times10^{-4}$. We did not conduct any extensive hyperparameter optimization, so more optimal configurations might exist. Table \ref{tab:parameters} shows the GPU memory usage during training, and it is observed that replacing the first layer of the models with an FDC layer does not affect memory usage. On the contrary, when building the entire model with FDC layers, GPU memory usage doubles, as does the number of parameters. It is worth noting that the training process varies between around \SI{3}{hours} and \SI{10}{days} for the base CNN architecture and the full FDC-CNN architecture, respectively. The evaluation, which includes the computation of the 32-channel SCMs for more than \SI{3}{hours} of audio, the model inference, and the RMSE computation, takes approximately \SI{1}{hour} and \SI{20}{minutes} for all architectures.

%To quantitatively assess model performance, we compute the root mean squared error (RMSE) between the predicted 32-channel complex covariance matrices and the corresponding ground-truth covariance matrices. To analyze the frequency-dependent behavior of the upsampling process, RMSE and random RMSE values are additionally reported across all frequencies, for each one of the proposed architectures, in Fig. \ref{fig:freq_rmse}. For qualitative evaluation, delay-and-sum beamforming is applied to all covariance matrices (input, ground truth, and predicted). Furthermore, for the computation of the beamforming the implemented transfer function on signals of a rigid sphere, is the function as it is computed in \footnote{https://github.com/polarch/Array-Response-Simulator/blob/master/simulateSphArray.m}. Finally, the produced heatmaps are overlaid on the corresponding video recordings to visually assess spatial reconstruction accuracy. 

%\FloatBarrier
\section{Results}\label{section: results}
\FloatBarrier
\begin{table}[tb]
\centering
\begin{threeparttable}
\caption{Number of parameters, peak GPU memory consumption during training and RMSE value for each model}
\centering
\begin{tabular}{
|>{\raggedright\arraybackslash}p{0.38\columnwidth}
|>{\raggedleft\arraybackslash}p{0.14\columnwidth}
|>{\centering\arraybackslash}p{0.14\columnwidth}
|>{\centering\arraybackslash}p{0.14\columnwidth}|
}
\toprule
\centering\textbf{Model architecture} & \centering\textbf{Parameters} & \textbf{GPU memory} & \textbf{RMSE} \\
\midrule
Random guess & & & 0.548\\
Base CNN & 6.29 M & \SI{2.03}{\giga\byte} & 0.452\\
%\hline
Expanded CNN & 191.29 M & \SI{4.81}{\giga\byte} & 0.433 \\
%\hline
Hybrid FDC-CNN Base & 6.69 M & \SI{2.04}{\giga\byte} & 0.447\\
%\hline
Hybrid FDC-CNN Expanded & 192.09 M & \SI{4.82}{\giga\byte} & 0.432\\
%\hline
%Hybrid 2FDC-CNN & 406.46 M & 0.448\\
%\hline
Full FDC-CNN & 539.36 M & \SI{10.13}{\giga\byte}  & 0.451\\
\bottomrule
\end{tabular}
\label{tab:parameters}
%\begin{tablenotes}
%\footnotesize
%    \item RMSE loss values is the RMSE value of the model averaged throughout all 32 microphones.
%\end{tablenotes}
\end{threeparttable}
\end{table}

\begin{figure}[tb]
    \centering
    \includegraphics[width=\linewidth, height=0.7\textheight,
  keepaspectratio]{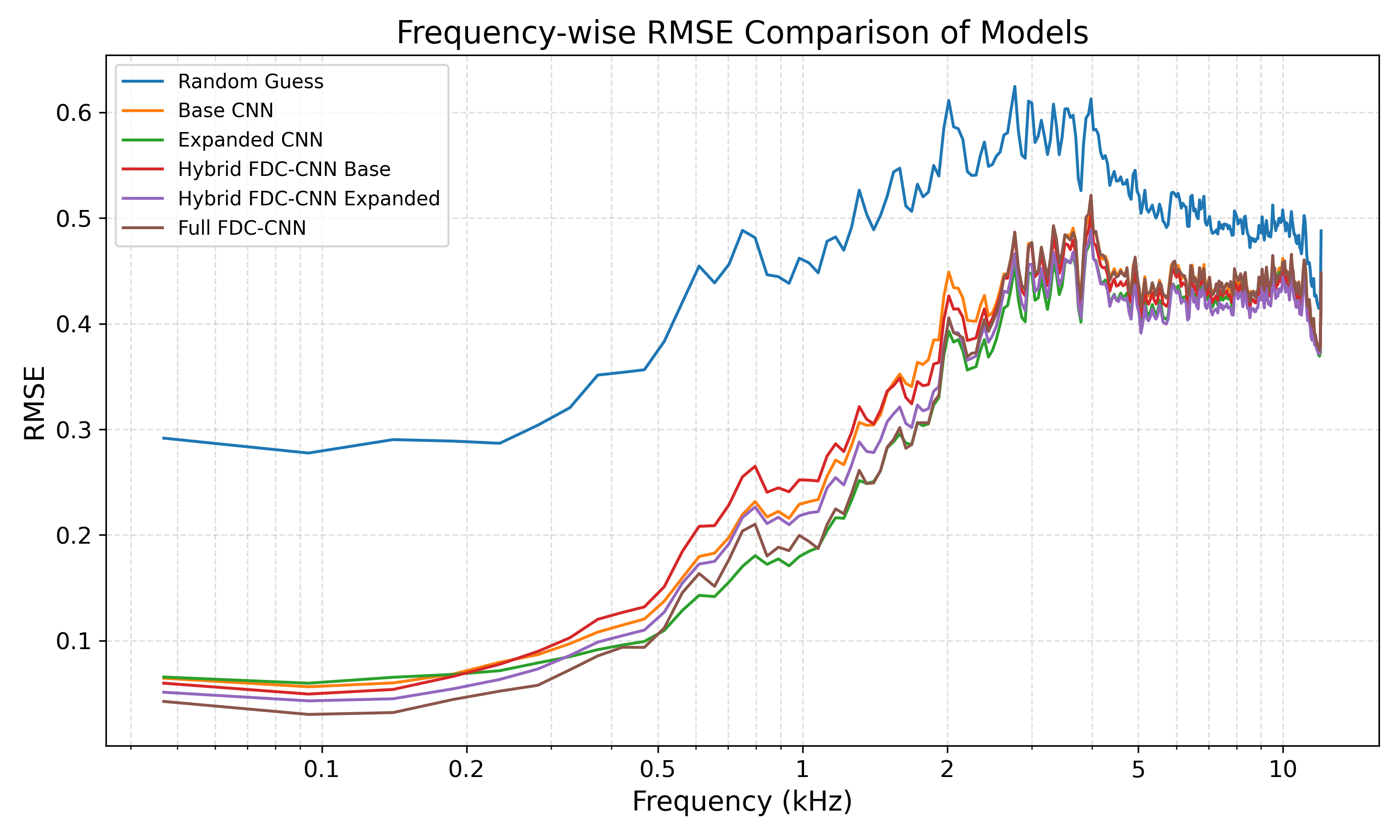}
    \caption{RMSE loss per frequency (log. scale) for each proposed model and the random guess.}
    \label{fig:freq_rmse}
\end{figure}
To quantitatively assess model performance, we compute the root mean squared error (RMSE) between the predicted 32-channel SCMs and the corresponding ground-truth SCMs on the STARSS23 dev-test dataset and average them across frequency. As a reference, we also include the RMSE of the random guess solution, which we define as the average $\mathbf{C}_x(t,f)$ of the dev-train groundtruths. As we can see in Table~\ref{tab:parameters}, there are not big differences between models, but they all clearly outperform the random guess, proving that they are not only exploiting dataset biases, but they are able to extract information from the 4-channel input SCM. 

The use of architectures with larger number of convolutional channels seems to have a bigger impact than the use of FDCs, though using FDCs at the first layer of the models provides some minor improvements of the performance with a negligible impact in the model size. Contrary to this, replacing all the convolutional layers by FDCs radically increases the model size and has a negative impact on the results, probably because of not having enough data to exploit higher power of a bigger model. % Furthermore, in Table \ref{tab:parameters}, for each proposed architecture is mentioned the GPU memory usage during training, where it is observable that replacing the first layer of the models with FDC layer is not impacting the use of memory during the training, on the contrary when building the whole model with FDC layers then the usage of GPU memory almost doubles without resulting in significantly better results. It is worth noting that the training process varies between $\sim 3~\mathrm{hours}$ for the base CNN architecture and $\sim 10~\mathrm{days}$ for the full FDC-CNN architecture and inference code runs approximately for $\sim 1~\mathrm{hour}$ and  $\sim 20~\mathrm{minutes}$ for all the proposed architectures.

To analyze the frequency-dependent behavior of the upsampling process, the RMSE of the SCMs of each frequency bin are additionally reported in Fig. \ref{fig:freq_rmse}. As could be expected, the estimation error grows with frequency till reaching the aliasing limit of the 4-channel array used as input for the model. However, even after exceeding that limit, the models are still able to provide better estimates than the random guess. This is probably because the models are able to use the information of the lower frequencies to predict the high frequency SCMs.

%The RMSE value is averaged over all the time-frequency bins of all the covariance channels. It can be seen that \textcolor{purple}{we have to write which is better and why}. Additionally, in Table \ref{tab:parameters} are presented the number of parameters of all proposed architectures, that each model require.

In Fig. \ref{fig:covmat}, three SCMs of a dataset sample, are presented. To the left it is shown the 4-channel input SCM, in the middle is shown  the ground truth 32-channel SCM,  and to the right is shown the 32-channel SCM predicted by the Hybrid FDC-CNN Expanded architecture; all of them are averaged across all frequency bins, for a specific time frame. We can see how the patterns present in the groundtruth seem to be intensified in the estimation of the model, which looks like a denoised and sharpened version.

%Fig. \ref{fig:covmat} illustrates the covariance matrix of the same segment as it is presented in Fig. \ref{fig:heatmap}.

%To qualitatively measure the covariance matrix upscaling, we also present in Figure \ref{fig:heatmap} the beamformed sound heatmaps overlaid on the video of the input 4 channel sparse microphone, upscaled 32 channel audio and the reference 32 channel audio. It can be clearly seen that our upscaling method localizes the active sound source robustly, although it struggles to model the reflections and reverberation (loosely worded, something on these lines). 

%We also investigate the performance of the upscaling model for all the frequency bins. It can be seen that from Figure \ref{fig:freq_rmse}, the models are able to upscale the covariance matrices accurately at lower frequencies and the performance decreases as frequencies get higher. (plausible reasons: )

%Furthermore, Fig. \ref{fig:freq_rmse} illustrates the RMSE values for all frequency bins for all the studied models. Therefore, it is observable that all architectures manage compute well the predicted values for low frequencies, but struggle to predict well the behavior for high frequencies. 

\begin{figure}[t]
    \centering
    \includegraphics[width=\linewidth, height=0.7\textheight,
  keepaspectratio]{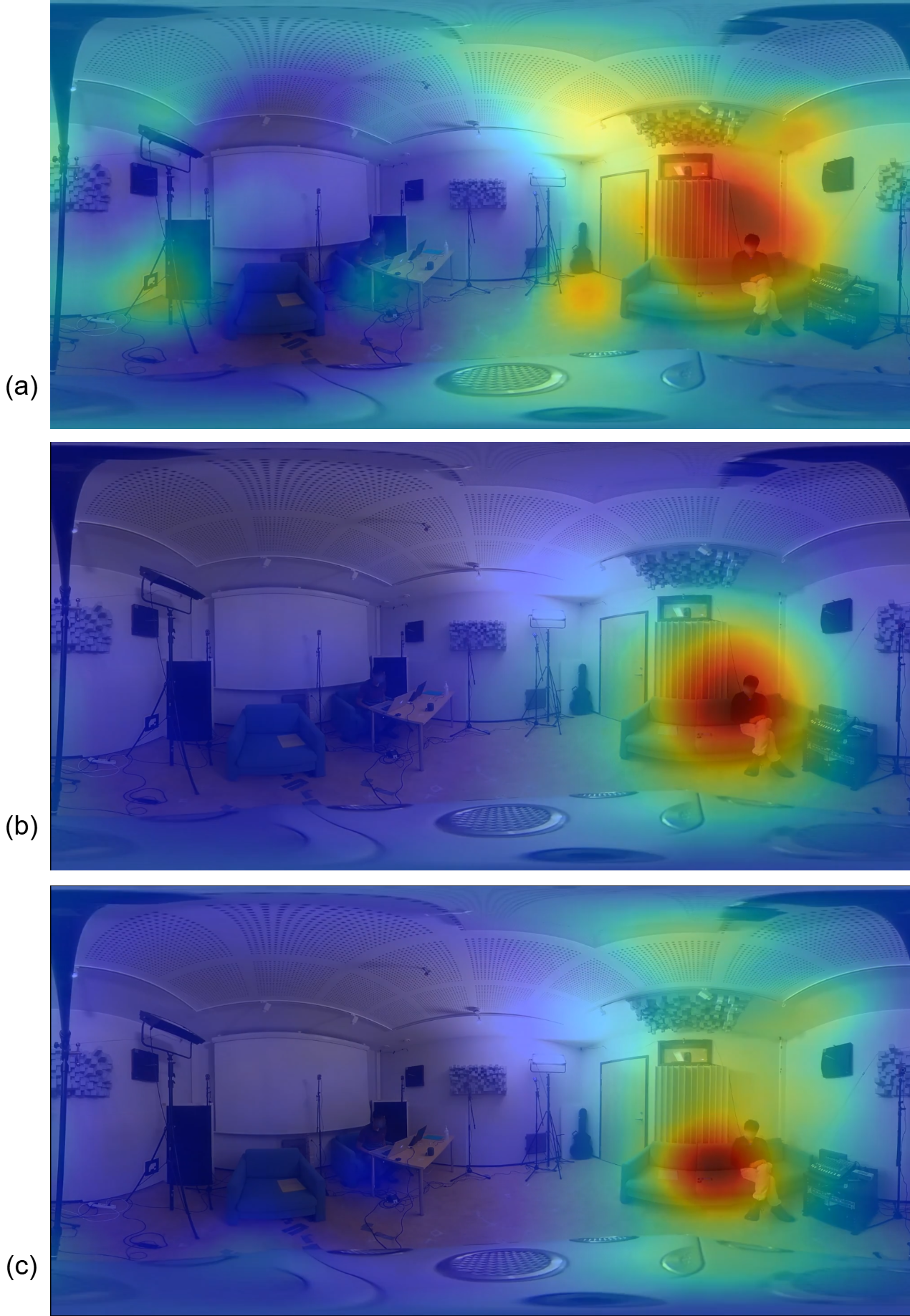}
    \caption{Beamformed sound heatmaps for (a) 4-channel, (b) predicted 32-channel and (c) reference 32-channel}
    \label{fig:heatmap}
\end{figure}

For qualitative evaluation, delay-and-sum beamforming is applied to the input, ground truth, and predicted SCMs. This computation is based on the analytical solution for the transfer function of a rigid sphere \footnote{\url{https://github.com/polarch/Array-Response-Simulator/blob/master/simulateSphArray.m}}. Finally, the produced heatmaps are overlaid on the corresponding video recordings to visually assess spatial reconstruction accuracy. Despite we focus on delay-and-sum beamforming, more advanced techniques could be employed to generate acousit images from the upsampled SCMs, including those based on deep learning such as DeepWave \cite{simeoni2019deepwave} and LAM \cite{LatentAcoustic}.

Fig. \ref{fig:heatmap} (a) presents the heatmap corresponding to the 4 channels input data, while Fig. \ref{fig:heatmap} (b) shows the heatmap of the predicted 32 channels and Fig. \ref{fig:heatmap} (c) depicts the heatmap for the ground truth 32 channels. In this example, the sound event corresponds to a ringing mobile phone and its spatial reflections. Therefore, it is worth to mention that the 32-channel original audio accurately localizes the sound source, whereas the predicted map is slightly shifted upwards, but it is still more accurate than the map from the original 4-channel input data. We can also observe how the heatmap generated from the estimated 32-channel SCM is more focused on the main audio source and ignores the ceiling and wall reflections present in the groundtruth. This is probably a consequence of the sharpening observed on the covariance matrices.

Despite being training to replicate the SCM of a 32-channel array without any kind of further processing, the models seem to focus on the stronger sources and filter out the wall reflections. This is probably due to the stronger sources being easier to find and model by the models, and could be an advantage or a disadvantage depending on the final application of the system.

\section{Conclusion}\label{section: conclusion}
This study compared five CNN model architectures for virtual microphone array upsampling from a tetrahedral 4-channel array to a spherical 32-channel, emphasizing  on sound event localization and spatial distribution. Performance evaluation  using RMSE and conventional beamforming highlights the challenge of direct upsampling from 4-channel covariance matrix to 32-channel, given the sparse spatial information. The Hybrid FDC-CNN Expanded architecture accomplished the lowest RMSE value of 0.432. All proposed models achieve better performance for low frequency over high frequency bins, due to aliasing limit of the 4-channels. However, each model consistently outperforms the random guess baseline at all frequencies, affirming its ability to learn spatial characteristics, successfully localizing sound events. In comparison to the ground truth, the predicted heatmaps are more concentrated to the sound sources while ignoring to model reflections and reverberations. The models with FDC layers  learn frequency dependent characteristics and with more modeling capacity, it performs better than the standard CNN models. Nevertheless, the optimal model choice  depends on application-specific requirements, taking into considerations the model complexity and frequency dependent characteristics, balancing performance and memory size of the implemented system. Future works will focus on reducing the computational complexity while maintaining or improving the spatial distribution on the produced acoustic maps.

\begin{comment}
\begin{figure}[h]
    \centering
    \includegraphics[width=\linewidth, height=0.7\textheight,
  keepaspectratio]{Block_diagram_4model.jpg}
    \caption{Example image}
    \label{fig:example}
\end{figure}
\end{comment}

%\section*{Acknowledgment}
%\newpage
%\FloatBarrier
\bibliographystyle{IEEEtran}
\bibliography{Eigenmik32_library_v2}

\end{document}